\documentclass[journal=nalefd,manuscript=letter]{achemso}
\usepackage[version=3]{mhchem} 

\usepackage{siunitx}
\usepackage{graphicx}
\usepackage{achemso}

\title{Effective out-of-plane g-factor in strained-\ce{Ge/SiGe} quantum dots}

\author{Andrew J. Miller}
\email{amille2@sandia.gov}

\author{Mitchell Brickson}

\author{Will J. Hardy}
\affiliation{Sandia National Laboratories}

\author{Chia-You Liu}

\author{Jiun-Yun Li}
\affiliation{National Taiwan University}

\author{Andrew Baczewski}
\affiliation{Sandia National Laboratories}

\author{Michael P. Lilly}

\author{Tzu-Ming Lu}
\affiliation{Sandia National Laboratories}
\affiliation{Center for Integrated Nanotechnologies}

\author{Dwight R. Luhman}
\affiliation{Sandia National Laboratories}

\begin{document}

\begin{abstract}
Recently, lithographic quantum dots in strained-\ce{Ge}/\ce{SiGe} have become a promising candidate for quantum computation, with a remarkably quick progression from demonstration of a quantum dot to qubit logic demonstrations.
Here we present a measurement of the out-of-plane $g$-factor for single-hole quantum dots in this material.
As this is a single-hole measurement, this is the first experimental result that avoids the strong orbital effects present in the out-of-plane configuration.
In addition to verifying the expected $g$-factor anisotropy between in-plane and out-of-plane magnetic ($B$)-fields, variations in the $g$-factor dependent on the occupation of the quantum dot are observed.
These results are in good agreement with calculations of the $g$-factor using the heavy- and light-hole spaces of the Luttinger Hamiltonian, especially the first two holes, showing a strong spin-orbit coupling and suggesting dramatic $g$-factor tunability through both the $B$-field and the charge state.
\end{abstract}

\maketitle


Semiconductor quantum dots have proven to be a promising platform for scalable spin-based qubit operations\cite{pet05,vel14,kim14,eng15,mau16,har17,zaj18,wat18,hua19}.
Much research has been performed using silicon-based electron quantum dots.
Holes in germanium exhibit many of the desirable properties of electrons in silicon, with some beneficial differences:
Absence of degenerate valley-states\cite{mor14}, the atomic $p$-orbital character that leads to a natural suppression hyperfine induced decoherence\cite{bul07,tes09,kea11,pre16}, and a large spin-orbit coupling\cite{mor14,cho18} that allows for control of qubits using electric dipole spin resonance\cite{bul07}.
Because of these potential advantages, strained-germanium quantum dots have been under investigation recently\cite{har19}.
Remarkably, hole-spin qubits in strained-\ce{Ge/SiGe} heterostructures have seen a quick progression from demonstration of a quantum dot to qubit logic\cite{hen20,hen20-3}.

The large separation between the heavy- and light-hole subbands in this material, which is enhanced by the strain, leads to a reduction in mixing between these subbands.
This results in reduced effective off-diagonal terms in a single-band model of the heavy-hole space.
Thus, the primarily heavy-hole character of holes in \ce{Ge/SiGe} quantum wells is expected to show large $g$-factor anisotropy between magnetic fields applied in-plane and out-of-plane\cite{nen03}.

This has been confirmed in one-dimensional channels\cite{miz18} and two-dimensional (2D) hole systems where the anisotropy was observed to be density dependent\cite{lu17}. 
For applications of single spin qubits in planar quantum dots, the magnitude of the anisotropy remains unclear.
The in-plane $g$-factor observed in single-hole qubit experiments is tunable with a value of $<\num{0.3}$\cite{hen20-2}.
Measurements of the $g$-factor of a single hole confined to a quantum dot with the magnetic ($B$)-field directed out-of-plane have not been reported. 
For many-hole quantum dots with an applied out-of-plane magnetic field, the bare $g$-factor can be obscured by orbital effects\cite{hof19}. 
A highly anisotropic $g$-factor introduces additional opportunities for optimization not present in other qubits. 
For example, the $B$-field needed to create sufficient Zeeman gap for spin readout would be significantly reduced for the direction with a large $g$-factor.
This may provide a trade-off for integration with superconducting elements, especially if the anisotropy is significant.

In this paper, we study the $g$-factor of a quantum dot in \ce{Ge/SiGe} down to the last hole using magnetospectroscopy with the $B$-field pointed out-of-the-plane of the sample.
We find an out-of-plane $g$-factor of \num{15.7} for a single hole.
Using the upper bound of the in-plane value reported in Ref.~\citenum{hen20-2}, $g_{\parallel}=\num{0.3}$, the anisotropy of the $g$-factor for a single hole confined to a lateral quantum dot is found to be over \num{50}. 
With increased hole occupancy in the quantum dot ($n>2$) we observe large changes in the apparent $g$-factor due to orbital effects.
In addition, we perform calculations using the Luttinger Hamiltonian with a realistic quantum dot potential and find remarkable agreement with the single hole $g$-factor and also capture the trend for higher occupancy up to six holes.

In this work, we used a strained-\ce{Ge} quantum well as our starting material.
The \ce{SiGe} barriers were composed of \SI{27}{\percent} \ce{Si}, with a \SI{62}{\nano\meter} barrier above the strained-\ce{Ge} well.
The strained-\ce{Ge} well was \SI{20}{\nano\meter} thick, and showed a mobility of \SI{1.84e5}{\centi\meter\squared\per\volt\per\second} in Hall bar measurements.
Details regarding the growth of the \ce{Ge/SiGe} heterostructure are reported in Ref.~\citenum{su17}.

\begin{figure}

	\centering
	\includegraphics[width = 0.5\columnwidth]{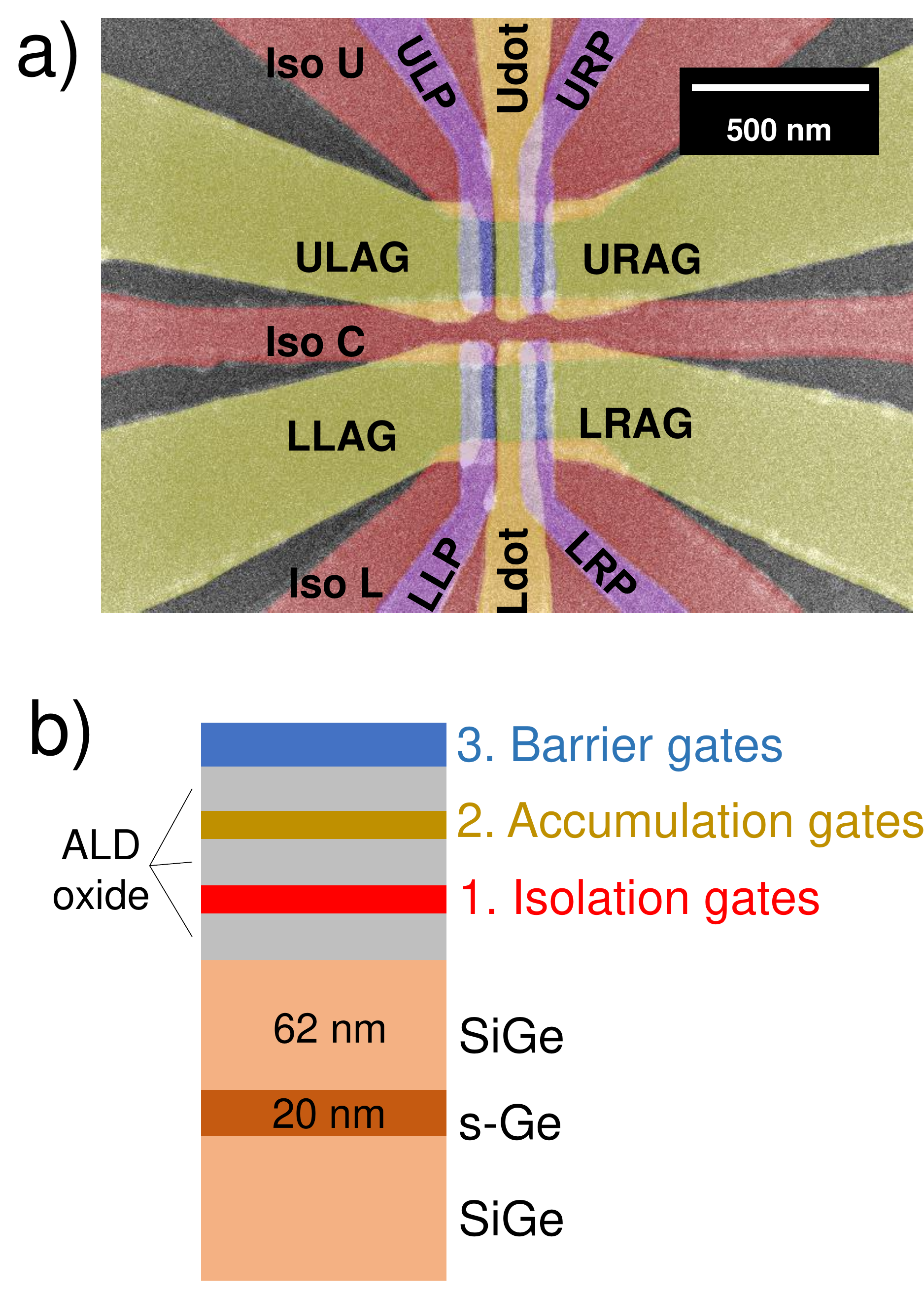}
	\caption{\label{fig:sem}
	SEM image of a device and the material stack.
	Subfigure a) shows an artificially colorized SEM image of the device design used in this work. 
	Subfigure b) presents the material stack, using the same color scheme as a).
	Charge carriers are acculumated in the strained-\ce{Ge} layer, and quantum dots are formed under the UDot and LDot electrodes, in the space between the isolation gates.
	}
\end{figure}

A lithographic quantum dot was patterned onto the strained-\ce{Ge} quantum well.
A scanning electron microscope (SEM) image of a device made using the same process as the measured device can be seen in Fig.~\ref{fig:sem}, along with a schematic of the material stack.
By manipulating the voltages of the patterned gate electrodes, charge-carriers (holes) are drawn into the strained-\ce{Ge} layer, producing a hole gas and hole quantum dots.

This electrode design defines two single quantum dots located next to each other on the device, labeled as upper and lower, each controlled by five gate electrodes.
Prior work leading up to this design can be found in Ref.~\citenum{har19-2}. 
Details regarding the device tuning and measurements can be found in the supporting information.

The data presented here were taken with an applied out-of-plane magnetic field in a dilution refrigerator with a base temperature of \SI{30}{\milli\kelvin}.
In this configuration, magnetospectroscopy measurements were made up to \SI{3}{\tesla}, as well as capacitance measurements between the upper dot and each electrode.
Through an evaluation of the thermal broadening of the charge sensing lines (see supporting information), the two-dimensional hole gas (2DHG) base temperature was found to be \SI{417\pm32}{\milli\kelvin}.

\begin{figure}
	\centering
	\includegraphics[width = 0.75\columnwidth]{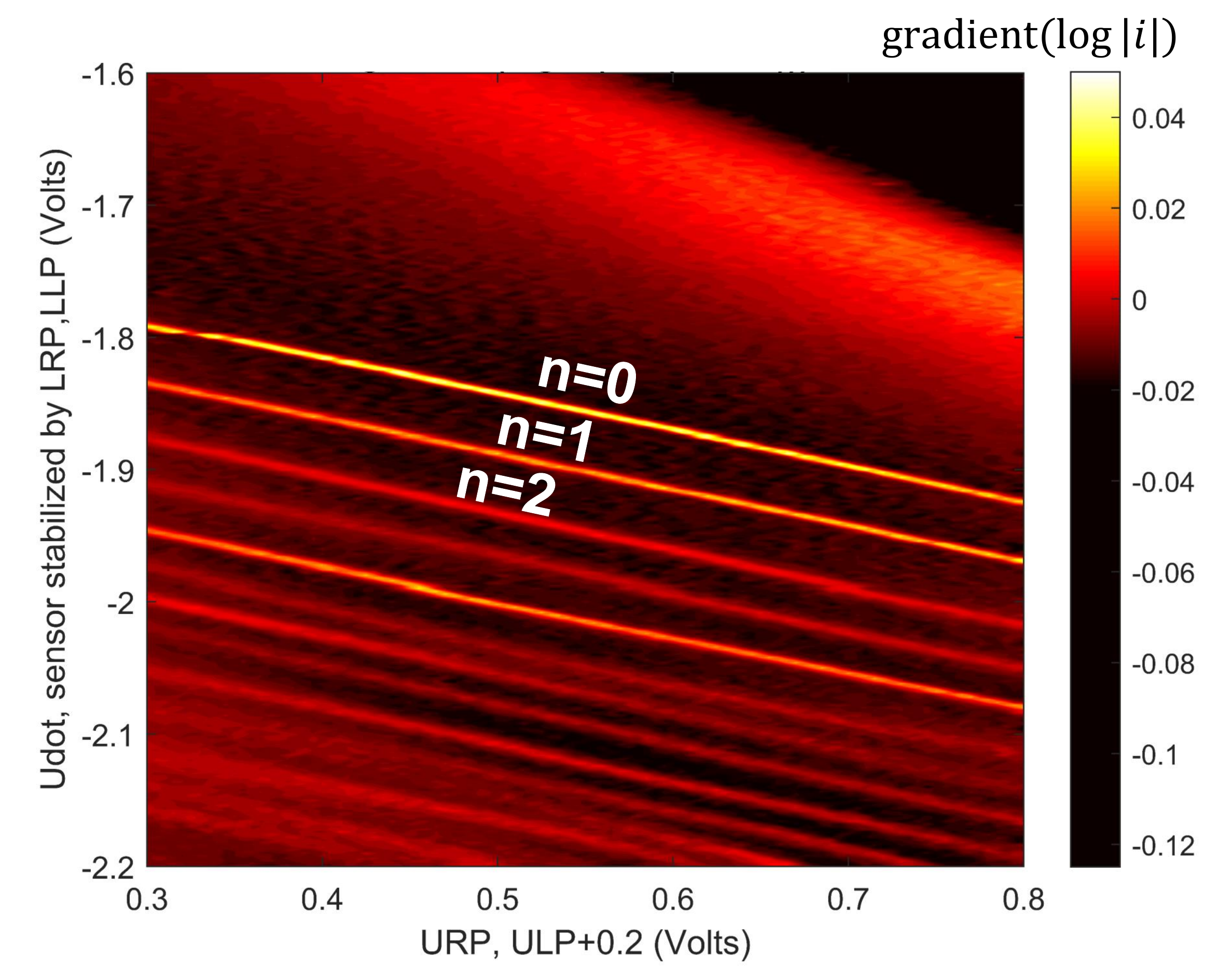}
	\caption{\label{fig:stab}
Stability plot of the quantum dot.
The voltage applied to the dot gate is shown on the vertical axis, while the barrier gates (plungers) are displayed horizontally.
Color indicates the gradient of the charge sensor current, $i$.
Horizontal labels indicate the voltage of URP at each point.
ULP was kept \SI{0.2}{\volt} lower than URP during this scan.
Details on this asymmetric potential can be sound in the supporting information.
The first three states of the quantum dot, with $n$ holes present in the dot are labeled.
}
\end{figure}

The lower dot was used as a charge sensor, while the upper dot was tuned to the single-hole regime. 
The coupling between the upper dot gate, Udot, and the quantum dot was measured to be $\alpha = \SI{75.4\pm3.5}{\milli\electronvolt\per\volt}$.
 Fig.~\ref{fig:stab} shows clearly resolved charge sensing lines.

\begin{figure*}
	\centering
	\includegraphics[width = 0.75\columnwidth]{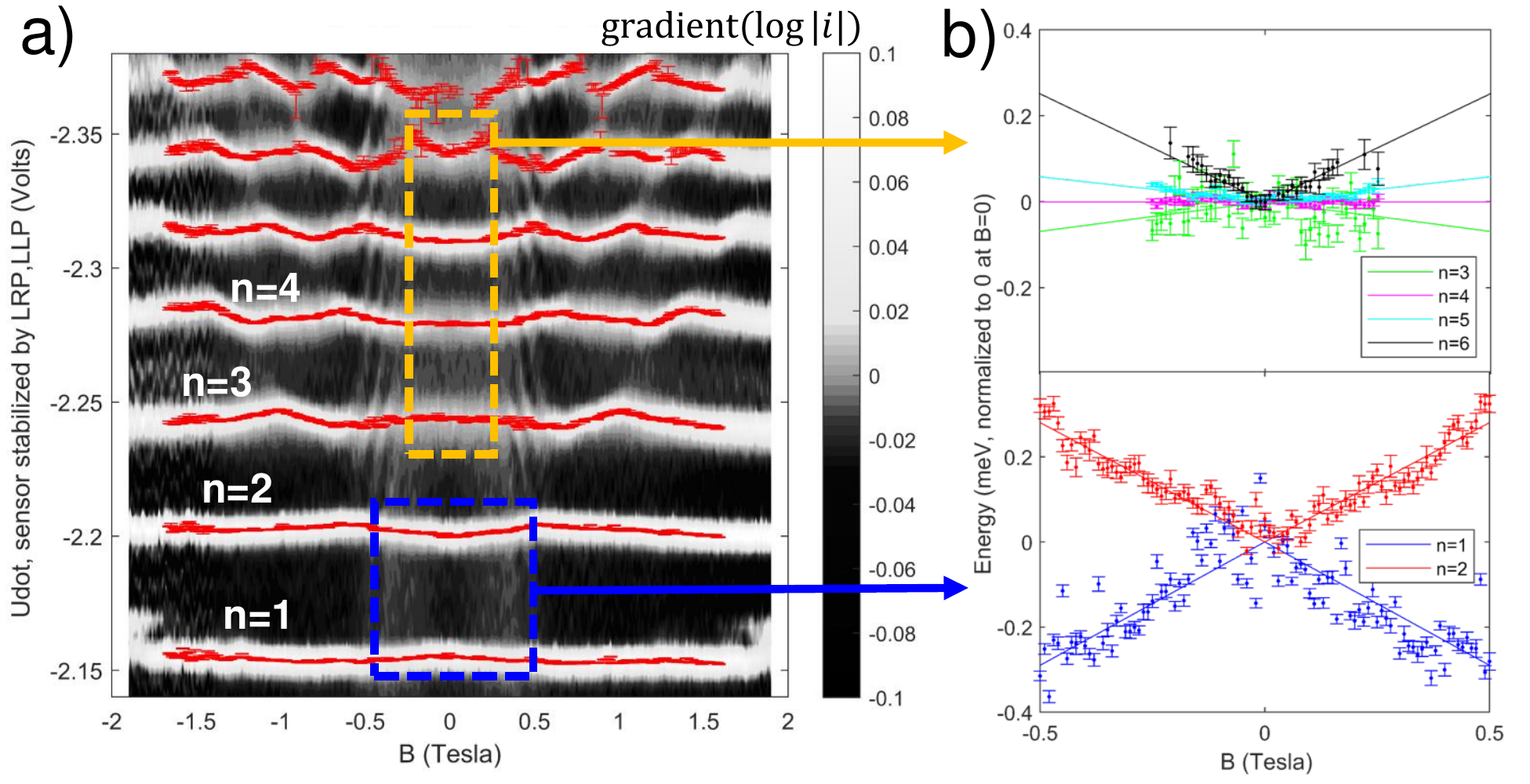}
	\caption{\label{fig:mag}
Magnetospectroscopy of the first several hole transitions.
Ten magnetospectroscopy scans such as the one shown on subfigure a) were taken, although not all scans were able to resolve as far beyond the first two lines.
This scan was taken using the Udot voltage to scan across states, while three of the ten scans were taken using the plungers.
Greyscale indicates the gradient of the charge sensor current, $i$.
Each line's position was extracted by fitting a Gaussian and is shown by the red overlay.
These points were converted to energy as a function of $B$, and the region near $B=0$ was fit using a linear, absolute value function, as shown in subfigure b).
Each line has been shifted in energy to 0 energy at $B=0$ in order to easily visualize the difference in slopes.
The lower plot shows the first two lines, from the $s$-like orbital, while the upper shows the next four lines, from the $p$-like orbitals.
The difference between the first and second line was used to extract the effective $g$-factor.
}
\end{figure*}

Magnetospectroscopy was used to determine the $g$-factor with the magnetic field oriented out-of-plane. 
Fig.~\ref{fig:mag} displays typical results.
Multiple data sets were taken by sweeping the Udot gate (as in Fig.~\ref{fig:mag}a) and also combinations of ULP and URP with Udot constant.
No significant differences were seen in the $g$-factors obtained by these two types of scans.

The data show various kinks as expected for orbital crossings in magnetospectroscopy. 
Additional near-vertical features near $B=\SI{\pm0.5}{\tesla}$ are also observed. 
These are likely the result of resonances with quantized states in the 2D reservoir due to the out-of-plane magnetic field.
Our primary concern here is determination of the $g$-factor near $B=\SI{0}{\tesla}$, away from these resonances.

Magnetospectroscopy data were analyzed by fitting Gaussians to trace each charge-transition line.
The maximum of each Gaussian fit is highlighted as a red overlay in Fig.~\ref{fig:mag}a.
Uncertainty in the position of each Gaussian was determined from the \SI{95}{\percent} confidence interval of the fit.
Gate voltage is converted to energy using $\alpha$ and set to zero at $B=\SI{0}{\tesla}$, as shown in Fig.~\ref{fig:mag}b.

The out-of-plane $g$-factor, $g_{\perp}$, is extracted from each data set for the range $B=$\SIrange{-0.5}{0.5}{\tesla} by fitting the difference of the magnetic field dependent energies, $\Delta E$  for the $n=1$ and $n=2$ lines to a linear function and relating the slope to the $g$-factor using $\Delta E = g \mu_B B$, where $\mu_B$ is the Bohr magneton. 
A final value of $g_{\perp}$ is determined through a weighted average of the results from each data set and the uncertainty in $g_{\perp}$ is calculated using the standard deviation among the values of $g_{\perp}$ from each of the \num{10} scans.
The results of each scan individually are displayed in the supporting information.
Using this method, we find $g_{\perp}=\num{15.7\pm2.2}$ for the lowest-lying quantum dot orbital in our device.

Long-term drift was intermittently observed throughout the time that the data were collected and the device required occasional voltage tuning adjustments from scan-to-scan. 
Any drift that may have occurred during a single scan is naturally taken into account in the difference $\Delta E$.
Small scan-to-scan variation in individual values of $g_\perp$ was random. 
There was no apparent correlation with tuning adjustments or whether Udot or a combination of ULP and URP were used to sweep through the dot occupancy levels.
This suggests that changes in electric field did not alter the observed value of $g_\perp$ in these experiments. 
Said differently, any change in $g_\perp$ due to electric field is not resolved within the uncertainty of our experiments.

These out-of-plane measurements yield the first experimental result of $g_\perp$ in lithographic, strained-germanium quantum dots in the single-hole regime.
The value of the in-plane $g$-factor in similar devices used for qubit demonstrations has been reported as $g_\parallel=\textrm{\numrange{0.2}{0.3}}$, depending on device tuning~\cite{hen20-2}.
Taking the upper bound on $g_\parallel$, and using our value for $g_\perp$ of \num{15.7} we find a $g$-factor anisotropy of $g_\perp/g_\parallel=52$, much greater than the anisotropy of $<\num{20}$ reported previously using the Kondo effect\cite{hof19}.
To our knowledge, this is the largest $g$-factor anisotropy observed among quantum dots that might be used as qubits. 
It is important to note that our measurement of $g_\perp$ was done in the $s$-orbital of the quantum dot and represents the bare Zeeman $g$-factor of a hole in a quantum dot with perpendicular applied field.

To further emphasize this, consider a similar analysis using holes in the $p$-orbital.
If one were to ignore orbital effects and extract the $g$-factor from the difference between the $n=2$ and $n=3$ states, the effective perpendicular $g$-factor is \num{-19.0\pm3}, and from the difference between $n=3$ and $n=4$, the effective $g$-factor is \num{4.8\pm4} (in the range of $B=$\SIrange{-0.2}{0.2}{\tesla}).
This highlights the strong influence of orbital effects on the effective $g$-factor, and is likely the source of the weaker $g$-factor anisotropy seen in previous transport results obtained with out-of-plane $B$-fields\cite{hof19}.
With measurements down to the last hole, these results show the expected dramatic anisotropy with respect to the in-plane $g$-factor, and also indicate a strong variability in the effective $g$-factor between different charge states.
This variation in effective $g$-factor as a function of charge state is also due in part to the strong spin-orbit coupling, an effect which has been seen in other hole-based qubits\cite{hen20, lil18, nen03}.

To better understand the mechanisms at play in the determination of $g_\perp$, calculations of the quantum dot energy spectrum were performed based on the heavy- and light-hole subbands of the Luttinger Hamiltonian\cite{winkler2003spin}.

\begin{figure*}
	\centering
	\includegraphics[width = 0.75\columnwidth]{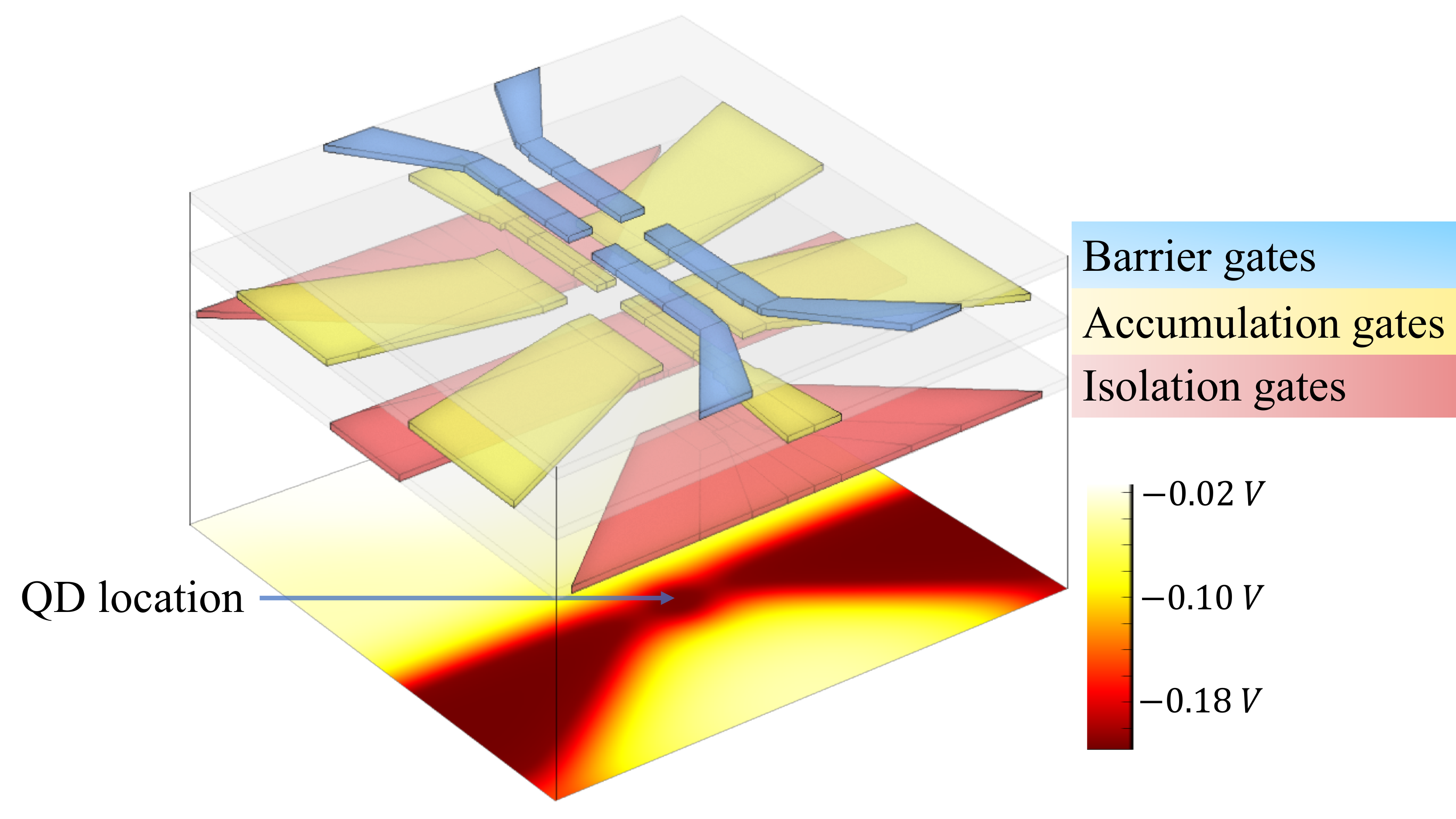}
	\caption{\label{fig:theory}
A wireframe design of the gate layout is included in a COMSOL model of the device.
This model allows an electrostatic simulation of the accumulation of the quantum dot and the charge in the leads using the Thomas-Fermi approximation for the charge density via a nonlinear Poisson solve.
The resulting potential is used as an input to our model of the electronic structure of our hole quantum dot to calculate the eigensates of the potential using an effective mass model.
}
\end{figure*}

The quantum dot potential was generated using a COMSOL Multiphysics\cite{comsol} calculation with the actual physical device geometry.
The potential was then further tuned to the experiment to match the crossing observed between the first and second excited states at $B=\SI{0.6}{\tesla}$ (Fig.~\ref{fig:mag}a) by adjusting the orbital gap to \SI{\approx 1.25}{\milli\electronvolt}.
This was accomplished by rescaling the potential in the x- and y-dimensions before solving for eigenstates without disturbing the potential in the z-direction.

We numerically study the $B$-field dependence of the spectrum of single-particle states for a realistic electrostatic model of the device to produce theoretical estimates for the $g$-factor.
Our framework first considers the light- and heavy-hole bands in isolation and solves for the spectrum of eigenstates of the associated effective mass Hamiltonians including the COMSOL-generated electrostatic potential (Fig.~\ref{fig:theory}).
We do so using an interior penalty discontinuous Galerkin discretization of the associated partial differential equations implemented in an in-house solver.
These numerically generated eigenstates were used as a basis for a Luttinger Hamiltonian, which explicitly treats the mixing of these bands due to spin-orbit coupling.
The impact of strain due to the lattice mismatch between the \ce{Ge} and \ce{SiGe} layers was added in the form of a Bir-Pikus Hamiltonian.
This leads to a splitting of the subbands in energy, in which the lower energy single-particle states are dominated by the heavy-hole band.

\begin{figure*}
	\centering
	\includegraphics[width = \columnwidth]{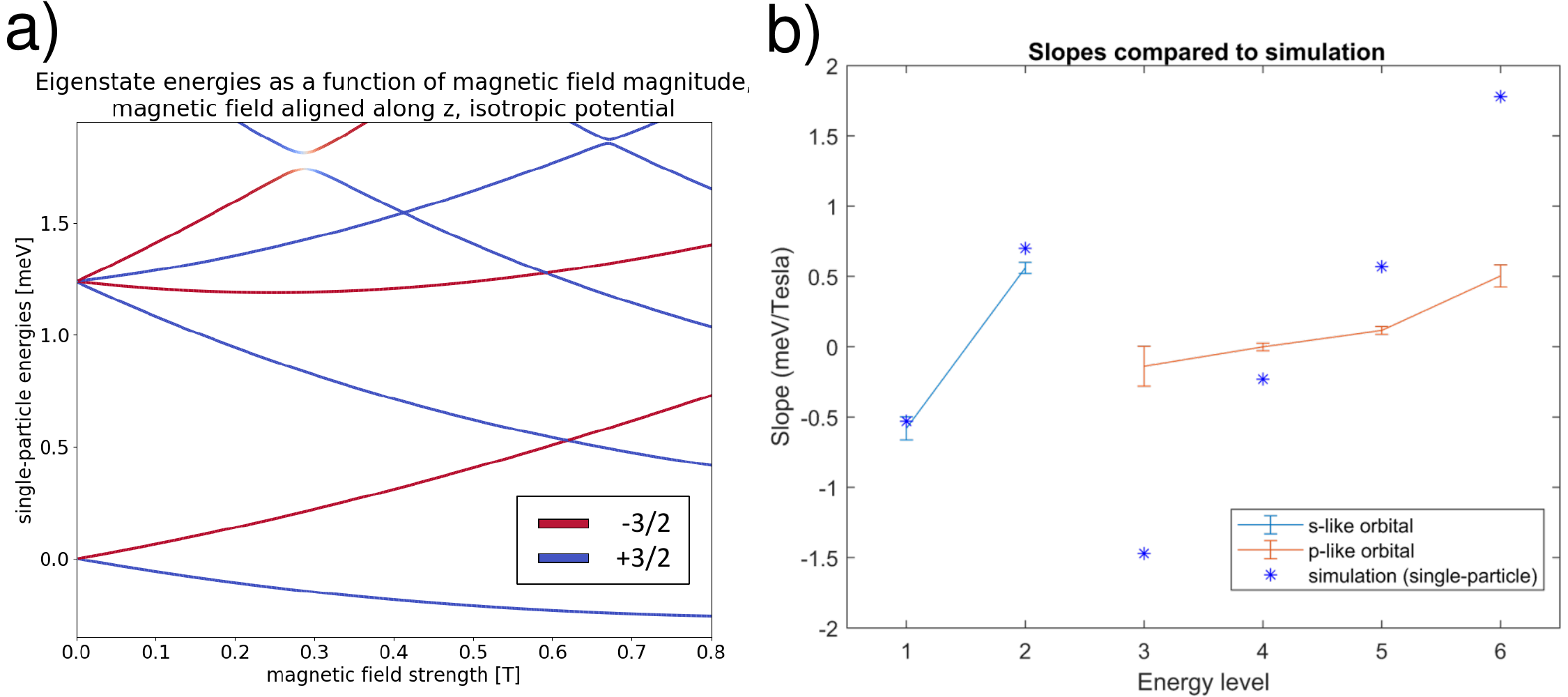}
	\caption{\label{fig:compare}
Theoretical calculation of the single-particle energy spectrum for a dot in a potential that is isotropic in the xy-plane shown in a).
This plot shows both the orbital and spin effects on the energy simultaneously.
The dominant spin eigenstate is shown by the indicated colors.
The splitting between the lowest two states is dominated by the Zeeman splitting, but the upper four states have both Zeeman and orbital effects that determine the slopes of the lines.
b) Comparison between the experimental and theoretical slopes of the energy levels with respect to magnetic field strength.
The simulation shows an excellent agreement regarding the first two states in the $s$-like orbital, however only presents the correct sign of the next 4 $p$-like states.
}
\end{figure*}

The energy spectrum from the simulation is shown in Fig.~\ref{fig:compare}a.
Due to the band gap in strained-germanium, these states are all spin \num{\pm3/2}. 
The average slope of each line from \SIrange{0}{0.2}{\tesla} was calculated and is compared with the slopes obtained from experiment using a linear fit in the range \SIrange{0}{0.2}{\tesla} in Fig.~\ref{fig:compare}b. 
Note that to emphasize the $s$- and $p$-orbital nature these are the slopes of individual lines, not the difference in lines used to calculate $g_\perp$ above. 
The experimental slopes shown are taken from the data in Fig.~\ref{fig:mag}b.

From these simulations, $g_\perp$ was calculated from the first two levels at \SI{0.2}{\tesla} using $\Delta E/ \mu_B B$ and found to be \num{21.25}, in reasonable agreement with the experimental result of \num{15.7\pm2.2}.
When compared to the $p$-orbital states, the simulations correctly reproduce the signs, but overstimate the magnitude.
This may suggest that the interactions between holes have a large effect, as those are not accounted for in this single-particle simulation.

General trends of the experimental data are reproduced by this simulation, including the signs of the slopes, as well as the ordering of their magnitudes.
The curvature of each state due to increasing magnetic confinement can also be seen in both the experiment and simulation.
One possible explanation for the reduced slopes seen in experiment may be due to leakage of the wavefunction into the \ce{SiGe} barriers.
The Luttinger parameter is much lower in \ce{Si} than \ce{Ge} ($|\kappa|\approx\num{0.42}$, compared to $\approx\num{3.41}$)\cite{winkler2003spin}.
It is expected that $|g| = 6|\kappa|$, and experimentally $g\approx2$ for holes in \ce{Si}\cite{voi16}.
This was suggested in Ref.~\citenum{hof19} as an explanation for the small value obtained for $g_\perp$.
Although our work suggests that orbital effects are a primary cause of their reduced effective $g$-factor, the $g$-factor in the \ce{SiGe} barriers may still influence the measurements presented here.

It is also important to note that although the slopes show repeated signs between states 3 and 4 as well as the states 5 and 6, the spin filling follows an alternating pattern, as shown by the color axis in Fig.~\ref{fig:compare}a.
These repeated signs are due to the $p$-like nature of these orbitals.
Angular dependence of the phase of the wavefunction in each orbital is aligned with or against classical cyclotron motion, causing the $B$ field to raise or lower the energy of each state.
Stated another way, the orbital angular momentum alignment (anti-alignment) with the $B$ field causes states 3,4 (5,6) to decrease (increase) with the increasing $B$-field, overpowering the Zeeman splitting due to the spin state.

Strained-germanium continues to stand out as a promising material for realizing qubits with lithographic quantum dots.
This experimental verification of the large $g$-factor anisotropy in the single-hole regime suggests that a very weak $B$-field can provide a sufficient Zeeman gap in the out-of-plane configuration.
Future designs may be able to take advantage of this fact, allowing integration with superconducting elements and the ability to perform spin readout using a $B$-field produced by on-chip components if desired.
These results also highlight the strong orbital-effects present in the out-of-plane case for $p$-like states.
While the simulations presented show good agreement in the single-hole case, they also highlight the complexity of the higher-order states.
These behaviors allow for dramatic tunability of the $g$-factor via the charge occupation and the $B$-field, and highlight the difference between $g$-factor measurements in the single-hole and many-hole regime.

\bibliography{refs}{}

\begin{acknowledgement}

This work was supported by the Laboratory Directed Research and Development Program at Sandia National Laboratories and was performed, in part, at the Center for Integrated Nanotechnologies, a U.S. DOE, Office of Basic Energy Sciences user facility.
Sandia National Laboratories is a multimission laboratory managed and operated by National Technology and Engineering Solutions of Sandia, LLC, a wholly owned subsidiary of Honeywell International Inc., for the U.S. Department of Energy’s National Nuclear Security Administration under contract DE-NA0003525.

This paper describes objective technical results and analysis. Any subjective views or opinions that might be expressed in the paper do not necessarily represent the views of the U.S. Department of Energy or the United States Government.

This work at National Taiwan University (NTU) has been supported by the Ministry of Science and Technology (109-2112-M-002-030-,109-2622-8-002-003-).

\end{acknowledgement}

\begin{suppinfo}
Additional details regarding the experimental methods and results are available.

\end{suppinfo}

\end{document}


\section{Experimental Methods}

\subsection{Measurements}
Measurements were performed in a commercial dilution refrigerator at $T\sim$ \SI{30}{\milli\kelvin}.
During this cooldown, a slight leakage current was seen on the upper-left accumulation gate (ULAG), preventing transport measurements through the upper side.
Because of this, the upper dot was tuned into the few-hole regime, while the lower dot was used as a charge sensor.
Typical voltages which were found to produce good results were \SIrange{-1}{-2}{\volt} on the lower accumulation gates, and near \SI{0}{\volt} on the isolation gates.
As a result of the leaky behavior on the ULAG during this cooldown, the upper accumulation gates had asymmetric voltages applied, only \SI{-0.9}{\volt} on the left, while the right was at \SI{-1.1}{\volt}.
Because of this asymmetry, the best tuning in the few-hole regime was found when the upper barrier, or ``plunger'', gates were operated with differing voltages also, with the upper-right plunger (URP) being set \SI{0.2}{\volt} above the upper-left plunger (ULP).
Plunger gate voltages of \SIrange{+1.75}{2.75}{\volt} were used on the lower, charge-sensing quantum dot, while voltages of \SIrange{+0.1}{+0.8}{\volt} were found to work well on the upper plungers, for the quantum dot with low occupation.
An ac signal with amplitude \SI{0.1}{mV} and frequency of \SI{607}{\hertz} was applied to the Ohmic contact on the charge sensing dot, and the resulting current through the charge sensor was measured using an SR830 lock-in amplifier.

\begin{figure}

	\centering
	\includegraphics[width = \columnwidth]{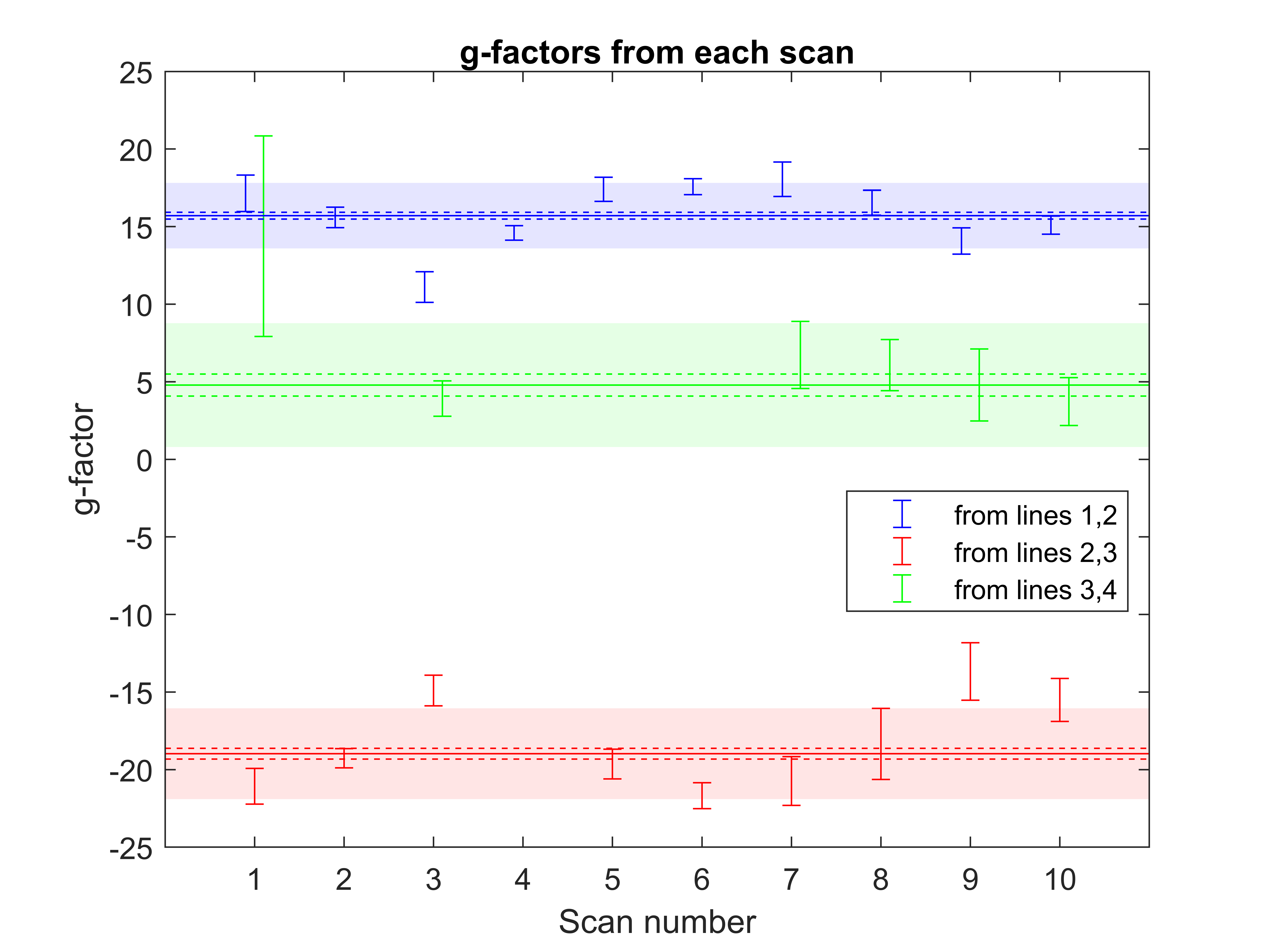}
	\caption{\label{fig:scans}
Effective $g$-factors obtained from each scan.
Scan number is shown on the horizontal axis, and corresponds to the chronological order in which they were taken.
Scans 4, 5, and 6 were taken by scanning URP, and using the relative capacitance between URP and Udot to determine the energy.
All other scans were performed by scanning Udot.
The three data sets show the results of the effective $g$-factor using different pairs of states.
Not all scans were able to resolve four lines, so some points are not present.
The solid lines indicate the weighted average of the results from each scan, and the dashed lines indicate the statistical uncertainty from this weighted average.
Standard deviation of each data set is indicated by the colored band and is the dominant source of the uncertainty we report.
}
\end{figure}

The resulting $g$-factor from each scan can be seen in Fig.~\ref{fig:scans}.

\subsection{Lever arm calculation}

\begin{figure}

	\centering
	\includegraphics[width = \columnwidth]{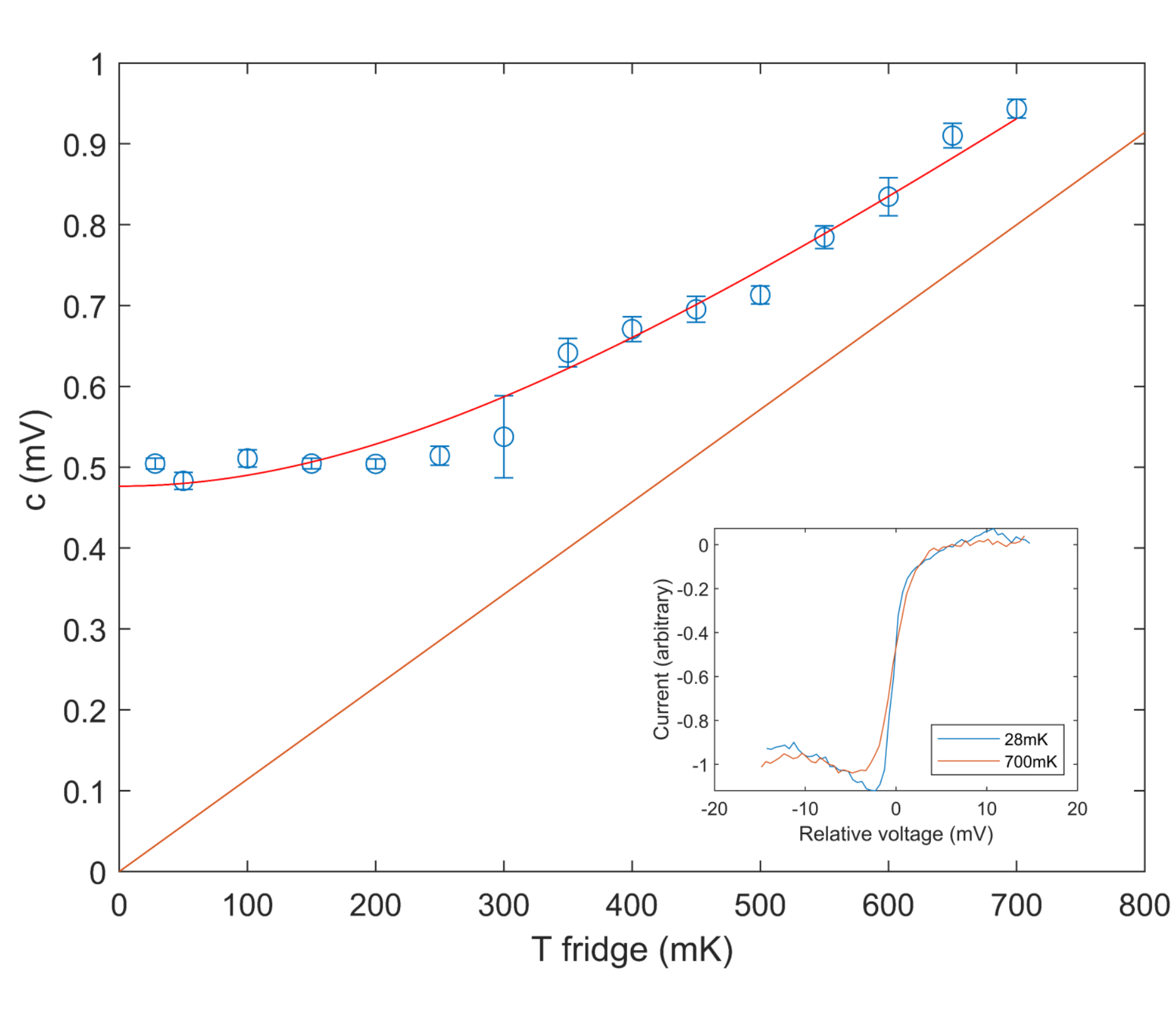}
	\caption{\label{fig:lever}
Line widths as a function of temperature.
The fitting parameter $c$ from the transition was used to extract the base temperature of the 2DHG and lever arm, ($\alpha=k_B / a$), of the dot electrode.
The inset plot shows an example of the transition width measured at two different temperatures.
To highlight the variation in width, the vertical axis of each transition has been normalized, and the horizontal axis has been shifted.
}
\end{figure}

To determine the lever arm of the dot electrode, the factor $\alpha$ representing the relation between the gate voltage and the potential of the dot, an analysis of the charge sensing transition vs. temperature was performed.
Due to the fact that one of the upper accumulation gates was not holding voltage well, it was not possible to determine the lever arm through the more traditional Coulomb diamonds, as transport could not be achieved through the upper dot.
With the upper dot in the single-hole regime, Udot was tuned to scan across the first charge sensing line.
This was scanned with the fridge held at a given temperature, and the current through the lower dot was measured.
Temperatures ranged from \SIrange{30}{700}{\milli\kelvin} with fifteen points being measured.

In each scan, the charge-sensing signal was fit with the expression
$$I=\frac{a}{\left( 1+\exp{\left[ (V-b)/c\right]}\right) }+d$$
where $I$ is the current signal through the charge-sensor, and $V$ is the scanning voltage applied to Udot.
From this equation, the fit parameter $c$ determines the width, and was plotted against the temperature, as shown in Fig.~\ref{fig:lever}.

From the relation to the Fermi-Dirac distribution, it can be seen that 
$$c = \frac{k_\textrm{B}}{\alpha}\sqrt{T_0^2+T_\textrm{bath}^2}$$
where $\alpha$ has units of \SI{}{\electronvolt}/\SI{}{\volt}, $T_0$ is the base temperature of the 2DHG, and $T_\textrm{bath}$ is the temperature the fridge was held at for the measurement.
From the fit shown in Fig.~\ref{fig:lever}, $T_0$ and $\alpha$ were extracted as \SI{417\pm32}{\milli\kelvin} and \SI{75.4\pm3.5}{\milli\electronvolt\per\volt} respectively.

\section{Beyond the fourth line}
\begin{figure}

	\centering
	\includegraphics[width = \columnwidth]{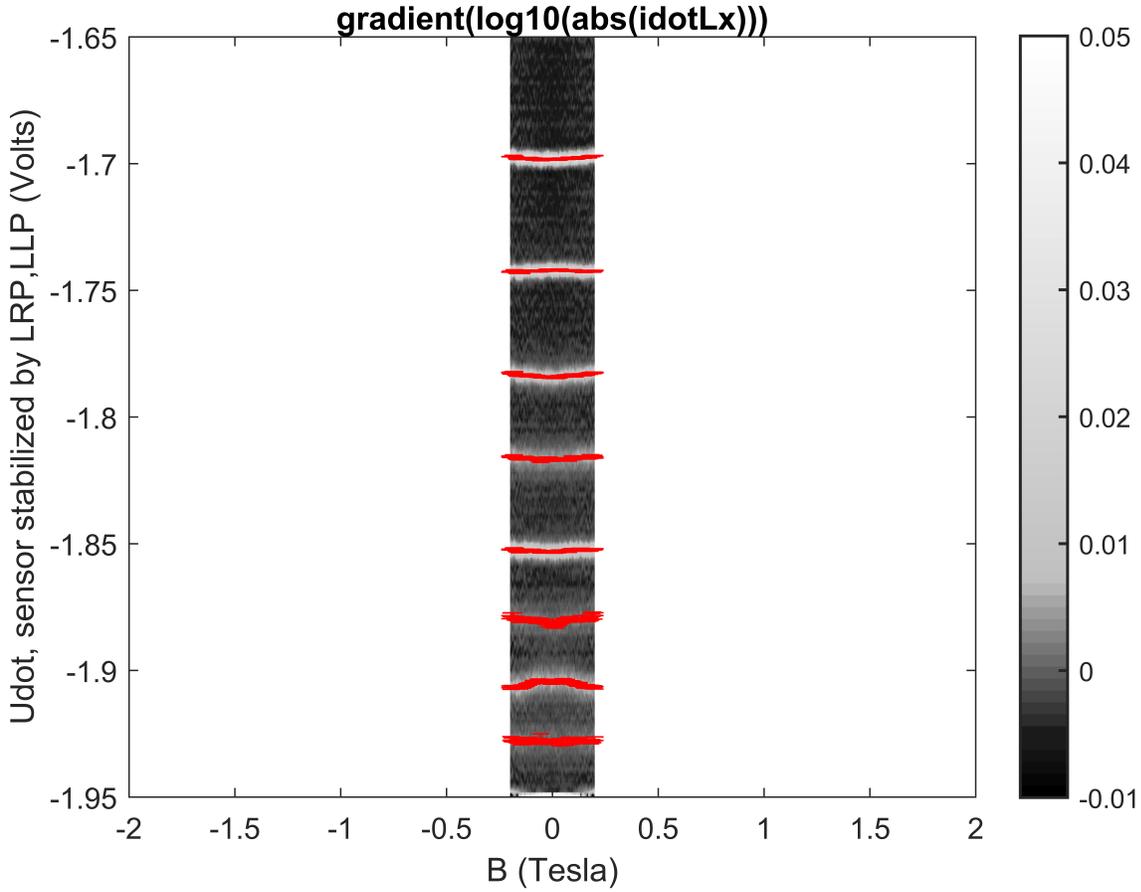}
	\caption{\label{fig:scan3}
Magnetospectroscopy data taken with an out-of-plane $B$-field.
This scan was unique in showing four lines with the same sign slope in a row.
All other scans were taken several days after this one, and showed the trend shown in the main text.
}
\end{figure}

Three scans were able to resolve magnetospectroscopy data up to the seventh line, however there was some discrepancy seen beyond the fourth line.
The first magnetospectroscopy scan showing up to the seventh line is shown in Fig.~\ref{fig:scan3}.
This scan was taken several days prior to the others, and showed lines with slopes following the pattern:+, -, +, +, +, +, -.
The scans taken later were consistent with each other, and showed slopes: +, -, +, +, -, -, +.

This discrepancy is likely due to the variations in tuning of the dot as it settled over the days between scans and/or a difference in static charges trapped at the interface.
With more holes present, the addition energy of each hole is reduced and variations in the ordering of excited states can easily occur with subtle changes.
Because of these observations, our analysis and comparison to simulation are primarily focused on the first four lines, which showed a consistent behavior during the experiment.